%% file: bare_conf.tex
\documentclass[10pt,conference]{IEEEtran}
\input{packages.tex}

\usepackage{cite}

\newcommand\intelxeon{Intel\textsuperscript{\textregistered} Xeon\textsuperscript{\textregistered} }
\newcommand\Orio{Orio}

\begin{document}
%
\title{Guiding Optimizations with Meliora:\\ A Deep Walk down Memory Lane}

\author{\IEEEauthorblockN{Kewen Meng, Boyana Norris}
\IEEEauthorblockA{Department of Computer and Information Science\\
University of Oregon\\
Eugene, Oregon\\
\{kewen, norris\}@cs.uoregon.edu}
}


%


\maketitle
\pagestyle{plain}

\input{abstract}


\IEEEpeerreviewmaketitle

\input{introduction.tex}
\input{background.tex}

\input{approach.tex}
\input{evaluation}

\input{related.tex}
\input{conclusion.tex}

\bibliographystyle{IEEEtran}
\bibliography{IEEEabrv,references}

\end{document}

%% file: packages.tex
\ifCLASSINFOpdf
   \usepackage[pdftex]{graphicx}
\else
\fi
\usepackage{amsmath}
\usepackage[ruled,vlined]{algorithm2e}

\makeatletter
\let\@ORGmakecaption\@makecaption
\long\def\@makecaption#1#2{\@ORGmakecaption{#1}{#2}\vskip\belowcaptionskip\relax}
\makeatother
\usepackage{listings}
\usepackage{url}

\usepackage{amsthm}
\makeatletter
\def\th@plain{%
  \thm@notefont{}
  \itshape 
}
\def\th@definition{%
  \thm@notefont{}
  \normalfont 
}
\makeatother

\newtheorem{definition}{Definition}

\usepackage{todonotes}
\usepackage{amssymb}
\usepackage{multicol}
\usepackage[caption=false]{subfig}
\usepackage{color,soul}
\usepackage{comment}

%% file: abstract.tex
\begin{abstract}
Performance models can be very useful for understanding the behavior of applications and hence can help guide design and optimization decisions. Unfortunately, performance modeling of nontrivial computations typically requires significant expertise and human effort. Moreover, even when performed by experts, it is necessarily limited in scope, accuracy, or both. However, since models are not typically available, programmers, compilers or autotuners cannot use them easily to guide optimizations and are limited to heuristic-based methods that potentially take a lot of time to perform unnecessary transformations. We believe that streamlining model generation and making it scalable (both in terms of human effort and code size) would enable dramatic improvements in compilation techniques, as well as manual optimization and autotuning. To that end, we are building the Meliora code analysis infrastructure for machine learning-based performance model generation of arbitrary codes based on static analysis of intermediate language representations. We demonstrate good accuracy in matching known codes and show how Meliora can be used to optimize new codes though reusing optimization knowledge, either manually or in conjunction with an autotuner. When autotuning, Meliora eliminates or dramatically reduces the empirical search space, while generally achieving competitive performance.
\end{abstract}

%% file: introduction.tex
\section{Introduction}
Performance models are structural representations of program behavior that can be used to describe and possibly predict application performance on one or more architectures. Such models provide software developers with useful information about potential bottlenecks and help guide them in identifying optimization opportunities. Models can also improve the quality of compiler optimizations or accelerate the process of empirical autotuning.

Traditional approaches to model-based performance code optimization employ a variety of techniques to represent code behavior of \emph{existing implementations}. Analytical models are more difficult to create but can also be used to represent the performance of \emph{potential implementations} that cannot be measured directly. Increasingly, machine learning (ML) methods are being used for performance modeling to identify patterns in the code feature space and identify optimal parameters. They are an attractive option because of the potential to capture complex patterns mostly automatically.

The main goal of our research is to provide a novel, scalable ML-based code matching methodology, and tool that enables accurate identification of loop-based computations. At present, when faced with a new code to optimize, most humans, compilers, and autotuners start from scratch or at best, apply some general heuristics when deciding on what code transformations to pursue. What we propose in this paper is a methodology to provide a high-quality starting point for the optimization of \emph{any computation}. When used by humans, this can save hours, days, or weeks of effort. When used in conjunction with compilers or autotuners, this approach can produce the performance that is competitive with that of empirically tuned codes but at a small fraction of the cost.

Our primary objective is to accelerate the code optimization process by using a deep learning technique to match a target code to similar computations that have been optimized previously. To accomplish this, we define a new graph-based code representation and combine it with a code generation framework to enable the automated creation of a deep learning model for matching loop-based computations. We name this approach Meliora, which translated from Latin means ``ever better''. The approach is based on learning accurate graph embeddings of a set of computational kernels ${K_0, K_1,..., K_N}$ that have been autotuned or manually optimized on the target architecture. When a new code $C_{new}$ must be considered, we apply the model to identify which optimized kernel, $K_i$, is the closest match. Based on that information and the autotuning results, we can select the best-performing version of $K_i$, $K_{i_{opt}}$, from our training set. This information can then be used by a human developer to manually optimize their implementation (which may involve significant refactoring), or it can be used by a compiler or an autotuner to automatically apply a small set of optimizations. The Meliora framework can thus greatly reduce or eliminate the exponential search space of potential optimizations.

\begin{figure*}[hbt]
  \centering
    \includegraphics[width=0.7\textwidth]{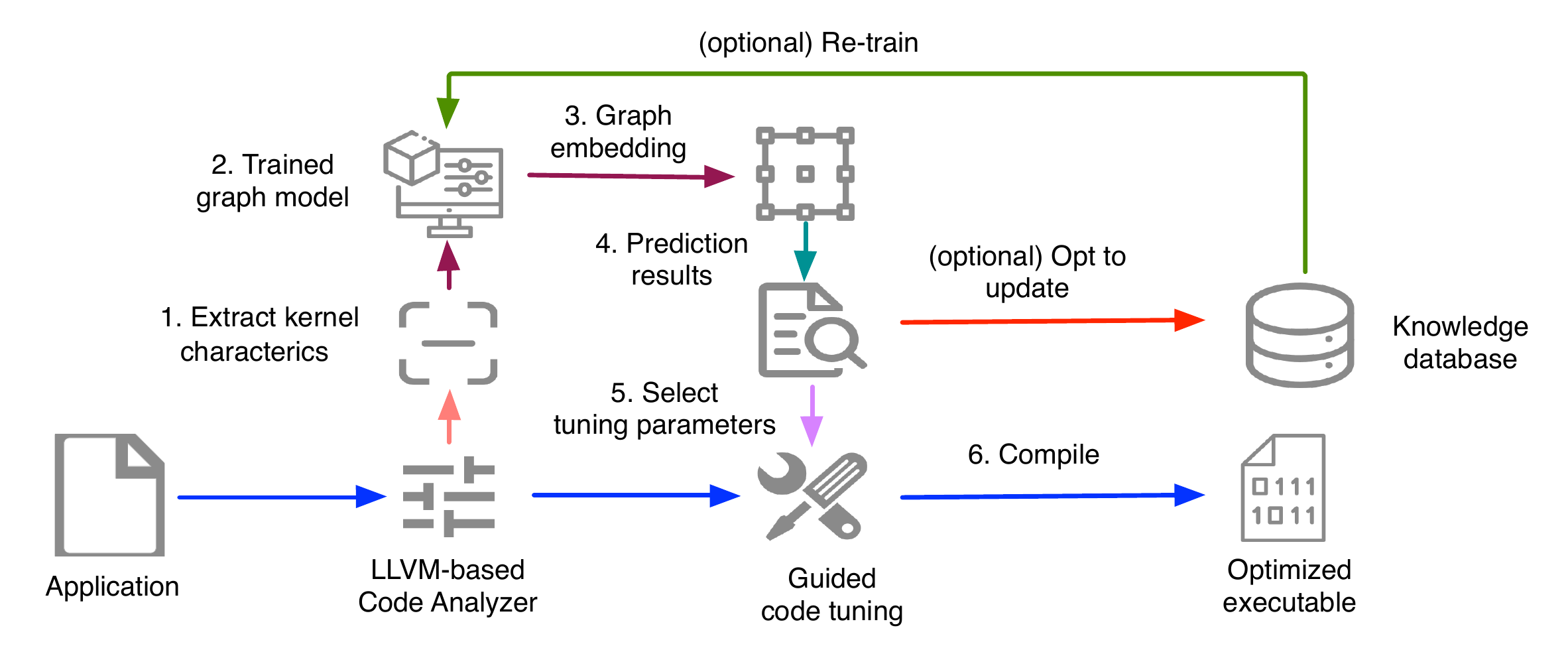}
  \caption{Meliora's workflow for code representation-based model generation and optimization.}
  \label{fig:workflow}
\end{figure*}

The two principal components of the Meliora approach are an LLVM-based frontend for extracting code features and an ML-based graph embedding component for learning efficient code representations. The overall workflow of feature extraction, model generation and subsequent code optimization is illustrated in Figure~\ref{fig:workflow}. 

Our contributions can be summarized as follows.
\begin{itemize}
    \item Definition of a graph-based code representation that extends the traditional control flow graph with computationally relevant features, such as instruction mix and reuse distance data.
    \item A deep-learning framework for learning code graph embeddings for efficient and accurate matching of computational kernels composed of code that does not contain non-standard function calls.
    \item Integration with an autotuner to enable fully automated construction of the training dataset for a supervised machine learning model used for matching the code graph embeddings.
    \item Evaluation of the accuracy and effectiveness of this approach on a set of computational kernels from the SPAPT benchmark suite.
\end{itemize}


%% file: background.tex
\section{background}\label{sec:background}
Before discussing Meliora in detail, we briefly overview the concepts and tools, on which we have based our approach.

\paragraph{LLVM}
Meliora's static analysis component is based on the LLVM compiler framework~\cite{lattner2004llvm}, which contains a set of open-source tools and libraries for writing code analysis and transformation tools for several languages and most HPC architectures. For example, the \textit{Clang} compiler frontend provides several useful static analysis and APIs to manipulate the Abstract Syntax Trees (ASTs) generated from C and C++ source code. \textit{LLDB} is the LLVM version of GDB for efficient debugging. \textit{Polly} applies the polyhedral model~\cite{pouchet.11.popl}, an intuitive parameterized algebraic representation, to optimize the memory access pattern within loops. The middle layer, LLVM intermediate representation (IR) connects the frontend and the backend of a program and, at the same time, offers a language-independent environment for developing new, portable tools, thereby reducing frontend development efforts. The metric generation components of Meliora (Sections~\ref{sec:cfg} and \ref{sec:analysis}) are hence built on top of the LLVM IR for compatibility with code written in different programming languages.

\paragraph{Graph representation learning}
Graph representation learning (or graph embedding) is a type of machine learning that focuses on creating compact vector representations of graphs. To be specific, for a given graph $G$ its goal is to find a mapping $f: v_i \to x_i \in \mathbb{R}^d $, such that the embedded vector $X_i = {x_1, x_2, ..., x_d}$ can represent the properties of the original graph. There are four general types of embeddings: node, edge, hybrid, and whole-graph embedding. Algorithms that operate on embeddings rather than graphs can greatly reduce the cost of computation and storage; hence embeddings are widely used in various types of applications, such as link prediction, node classification in social networks, and DNA analysis in computational biology. The challenge is to balance the compactness with the preservation of key properties. 
\textit{Random walk} is utilized by graph embedding approaches like DeepWalk~\cite{perozzi2014deepwalk} and Node2Vec~\cite{grover2016node2vec}, which samples a graph with many paths to find the context of the connected vertices. More recently, researchers have also employed convolutional neural networks (CNNs)~\cite{lecun1998gradient, lecun2015deep} and recurrent neural networks (RNNs)~\cite{mikolov2010recurrent} to learn graph representations, for example, in GraphSAGE~\cite{hamilton2017inductive}. We take a CNN-based approach to support code graph embedding because it is less limited than earlier approaches (e.g., we must handle directed graphs with node and edge labels).

\paragraph{Performance optimization} One of the common approaches to optimizing performance is automatic performance tuning (autotuning). Here we consider autotuning that not only explores simple parameters but rather considers significantly different code variants (e.g., through transformations such as tiling or automatic parallelization). \Orio~\cite{hartono2009annotation}, the autotuning framework we used to build the training dataset is based on annotated C or Fortran code, coupled with a tuning specification containing various transformations and their parameters. In general, Meliora does not require the use of an autotuner, however, it enables the relatively easy generation of a sizeable training dataset for our model.


%% file: approach.tex
\section{Methodology}\label{sec:approach}
Meliora is a novel framework for characterizing computational kernels whose goal is to dramatically reduce the time and effort required for optimizing performance. The primary objective is to accelerate the process of searching the space of optimizations by using a CNN-based technique to identify previously optimized similar codes. When used in conjunction with an autotuner, Meliora can greatly reduce or eliminate the exponential search space of parameterized code versions. Here we refer to a \emph{kernel} as any small to medium-sized computation consisting primarily of loops. The performance of many HPC applications is heavily dependent on the performance of a few key kernels, which would be the target for our analysis and optimization efforts. Unlike a library, Meliora does not aim to create a repository of ready-to-use functions optimized for particular architectures; rather, it provides a mechanism to \emph{discover} successful optimization of \emph{similar} (but rarely identical) computations. Also unlike library approaches, there is no specific limitation to the types of computation that can be considered.

The Meliora framework consists of two major components: front end for data collection and back end for data analysis. Figure~\ref{fig:workflow} shows the overall process of performing the front-end analysis to extract the code representation (step 1, Sec.~\ref{subsec:cfg}) and the data analysis (steps 2--4, Sec.~\ref{sec:learn}--\ref{sec:usingmodel}) in the backend.


\subsection{The Hybrid Control Flow Graph}\label{subsec:cfg}
The first step in extracting a code representation in Meliora is based on the traditional control-flow graph analysis.
A control flow graph (CFG) consists of nodes and edges describing all the possible execution paths of a program. 
The traditional CFG only contains nodes and edges that can provide limited information, such as the number of basic blocks and their connectivity. We can easily envision two codes with identical CFGs, but vastly different computations within basic blocks. For the purpose of precisely describing the structure and potential runtime behavior of a kernel, we require more information; hence, we introduce the hybrid CFG.

\begin{definition}[hybrid Control Flow Graph (hCFG)]
A directed graph denoted as $G = \langle V,E,\delta,\epsilon \rangle$ where vertex $V$ and edge set $E \subseteq V \times V $ stand for basic blocks and directed edges which connect them. In feature sets $\delta$ and $\epsilon$, $\delta_i(v_n)$ represents the information attached to node $v_n$ and $\epsilon_j(e_{mn})$ indicates the features attached to the edge from node $v_m$ to node $v_n$. 
\end{definition}

\begin{figure}[h!tb]
  \centering
    \includegraphics[width=0.8\linewidth]{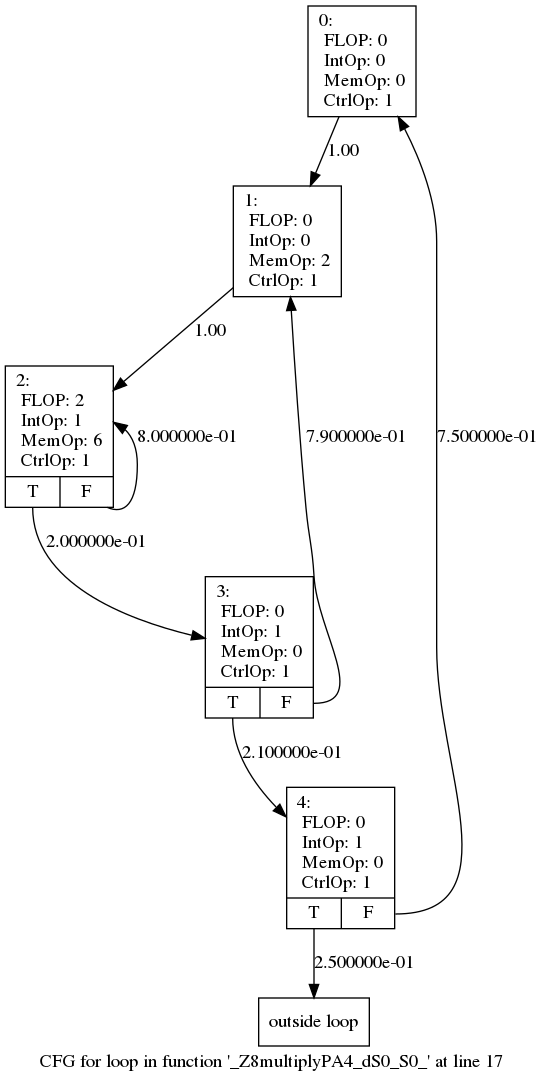}
  \caption{hCFG for the code fragment in List~\ref{list:1} (Meliora-generated graph).}
  \label{fig:cfg}
\end{figure}

The node and edge attributes are used for learning a representative model in the data analysis phase, which differentiates hCFG from regular CFG. Figure~\ref{fig:cfg} shows the generated hCFG for a matrix-matrix multiplication kernel as Listing~\ref{list:1}.

\begin{lstlisting}[caption={Matrix-matrix multiplication}, frame=bt, belowcaptionskip=0.8em, label={list:1}, basicstyle=\ttfamily\small]
    void multiply (double mat1[][N], 
                   double mat2[][N], 
                   double res[][N]) 
    { 
        int i, j, k; 
        for (i = 0; i < N; i++) { 
            for (j = 0; j < N; j++) {
                res[i][j] = 0.0; 
                for (k = 0; k < N; k++) 
                    res[i][j] += mat1[i][k] *  
                             mat2[k][j]; 
            } 
        } 
    } 
\end{lstlisting}

The graph structure of the hCFG is the same as that of the regular CFG, in which each vertex represents a basic block including a sequence of operations and each edge indicates the direction of execution flow. In addition, hCFG node attributes include a vector with instruction mix information. Edges are also annotated with attributes described later in more detail. To describe each node, we employ an instruction mix that consists of aggregated instruction counts of four major groups: floating-point, integer, memory access, and control operations. The node attributes categorize every node into different functional clusters by their dominating operation. For example, a floating-point-intensive node is one which contains a large portion of floating-point instructions. Similarly, a node with a majority of memory access instruction would tend to be memory-bound at runtime. 

Next, we compute the transition probabilities of each node (by using the method described in~\cite{lim2018similarity}) and attach the probability as an edge attribute.  Dynamic graph attributes distinguish our hCFG from the attributed CFGs used in security research~\cite{feng2016scalable}. They characterize the dynamic behavior of a kernel by lightweight runtime profiling, which offers insightful information about the propagation of values of interest but is not concerned with performance-relevant attributes. In Table~\ref{tab:1} we list the selected node attributes (shown as LLVM IR type) and edge attributes, and we describe the process of extracting these attributes in the following sections.

\begin{table}[h!tb]
\centering
\caption{Graph attributes in hCFG}
\label{tab:1}
{\renewcommand{\arraystretch}{1.5}
\begin{tabular}{|c|c|l|}
\hline
\multicolumn{1}{|c|}{\textbf{Type}}     
    & \multicolumn{1}{|c|}{\textbf{Attach to}}         
        & \multicolumn{1}{|c|}{\textbf{Source}}    \\ \hline\hline
         &                   & Floating-point: FAdd, FMul, FDiv, FSub  \\ \cline{3-3} 
Static   & Node              & Integer: Add, Mul, UDiv, SDiv, Sub   \\ \cline{3-3} 
         &                   & Memory: Store, Load, Fence, GetElementPtr \\ \cline{3-3}  
         &                   & Control: Ret, Br, Switch, Resume  \\ \cline{3-3}
         &                   & Memory Access estimation \\ \cline{3-3} \hline
Dynamic /  & Edge              & Transition probability: runtime profiling / \\
Static        &       ~                  &  Statically derived from IR \\  \hline 
\end{tabular}}
\end{table}

\subsection{hCFG metrics}\label{sec:cfg}
The metrics added to the hCFG directly affect the performance of the entire Meliora workflow and the accuracy of the generated model reflecting over the prediction results. For the selection of appropriate metrics as model features, we consider three aspects:  

\begin{enumerate}
    {\itshape\item Representability}: The metrics should be representative to describe and differentiate a program (or part of a program) which is of importance for feature learning. Moreover, a small number of representative metrics can greatly reduce the costs of model optimization.
    {\itshape\item Generation complexity}: As a lightweight modeling tool, one goal of Meliora is to provide efficient tuning guidance. Therefore on top of the above requirement, we also consider the complexity of metrics generation in terms of time and resource consumption.   
    {\itshape\item Integration compatibility}: Meliora can work with autotuners such as Orio~\cite{Norris:2007,hartono2009annotation} 
    in postmortem mode and also can behave as a standalone compiler. Thus, it requires the availability of all the metrics during compile time, so that Meliora can extract the metrics without requiring application execution.
\end{enumerate}

Table~\ref{tab:1} lists all the metrics required for building the hCFG and generating models based on the previous criteria. The rest of this section will discuss the metrics in detail.

\textbf{Instruction mix} is the collection of instructions counters which fall into four categories: Floating-point, Integer, Memory, and Control instructions. The instructions represent both high-level application information and the architecture on which the code will run. If we are looking for ways to accurately characterize each basic block, instructions are a natural choice. The Mira framework~\cite{meng2017mira}, for example, demonstrated the applicability and precision of the collection of instructions by static analysis. 

The type of a basic block is dominated by the majority category of its instructions. Accordingly, the nodes in hCFG exhibit different behaviors, for instance, some nodes are memory-intensive (e.g., a data structure initialization loop), while others are dominated by computation (e.g., matrix-matrix product). Therefore, instead of storing all instructions, we can compress the data attached to each node to capture just the four categories of instructions and thus further optimize the performance of model generation. Besides, by counting instructions, we are able to derive other valuable metrics, for example, arithmetic intensity of floating-point operations, to help researchers gain a deeper understanding of their code. All the stated reasons make the instruction mix an ideal metric for Meliora to extract and use to generate models.

\textbf{Memory access estimation}
In addition to the instruction mix, Meliora also generates reuse distance histograms to abstract the pattern of memory access within a basic block. Previous research~\cite{ding2003predicting, beyls2001reuse} shows that the reuse distance is an effective representation of data access locality and for further understanding of the behavior of memory hierarchy. This metric can be seen as a supplement to the instruction mix. This increases the dimension of the node attributes in the hCFG by the size of the reuse distance histogram (customizable, with default size 5), but it enables better differentiation between basic blocks than a simple instruction mix attribute.

The generation of the memory access estimation, however, is challenging for static tools. Most previous research on reuse distance analysis rely on the program instrumentation techniques as they are able to obtain the dynamic information from executions, such as memory references and instruction tracing, etc. Although such data may not be available in compile time, we can offer the reasonably estimated metrics derived from static program analysis, for instance, the range, upper (lower) bounds. In Meliora the memory access pattern is represented by a histogram vector whose number and size of bins are fully customizable by users. We describe the details of the implementation in the next section.

\textbf{Transition probability}
Besides vertex-attached metrics, the transition probability is used as the edge weight which indicates the probability of the execution flow from one basic block to another~\cite{lim2018similarity}. In real environments, it is not usually possible for a program to traverse the execution paths equally especially for multiple or deeply-nested branches. For example, in the code shown in List~\ref{list:1} certain basic blocks are executed more often than others depending on the loop static control part (SCoP). Consequently, it would affect the accuracy of the model if all execution paths were treated as equally likely. Therefore transition probability can be utilized to enhance the structural representability of the hCFG in which the frequent paths carry larger values than the infrequent ones. 

In Meliora the transition probability is the only metric that can be extracted through either static or dynamic analysis. As the unique edge attribute for the graph, it is of great importance, thus the data obtained from runtime profiling can describe the structure with emphasis on precision. However, in many cases generating the transition probability statically is sufficiently accurate and enables the entire Meliora workflow to run in pure static mode, which reduces the time for both creating and using the models.

\subsection{Metric extraction}\label{sec:analysis}
As we mentioned in Sec.~\ref{subsec:cfg}, all graph features can be computed statically and some (transition probabilities) can be optionally computed dynamically. In this section, we describe the static and dynamic techniques used in Meliora for obtaining the features and building the hCFG.

\textbf{Loops}
Loops are common and significant code structures in high-performance computing applications. Most computationally intensive code is expressed in loops and they account for the majority of application performance hotspots. In the paper by Bastoul et al.~\cite{bastoul2003putting}, the authors surveyed several high-performance applications. Their work shows that the average loop coverage, calculated as the ratio of in-loop statements to the total number of statements, ranges from 77\% to 100\% with an average above 80\% for the surveyed applications. 

The critical role and the characteristics of a loop make it the ideal target for code optimization. For instance, an optimizer (human, compiler, or autotuner) can apply several techniques such as loop fusion, loop unrolling, and loop tilting to reduce the cache miss rate by the enhancement of locality of the memory system in order to improve the overall performance of the generated code. Hence, Meliora focuses on the loops within kernels for extracting metrics and metrics and assisting in discovering code optimizations.

\textbf{Analysis granularity}
Kernels frequently contain more than one same-level loop. Each of these loops may have unique characteristics that would require a different approach to optimization. In other words, a single optimization strategy for the whole multi-loop kernel is unlikely to yield the best results. Hence, Meliora supports model generation based on metrics obtained from each top-level loop (or nested loops), so that later Meliora can generate the model and guide the optimization matching the same granularity in new kernels. In addition to the loop-basis analysis, we also provide users with the option for performing the static analysis in a coarser-grained manner, for example running on the function level with all loops combined. The goal of offering a high-level perspective is to further help the users to gain an understanding of the functionality of the kernels as a whole. If a kernel-level match is found, for example, that can help optimize larger portions of the code at once, thereby saving time and effort. 

\begin{algorithm}\label{algo:1}
\SetAlgoLined
\KwData{hCFG $G$}
\KwResult{vector $v$ as the histogram}
 current byte tracker $bt = 0$\;
 \While{$G$ has basic block ${BB}_k$ not visited}{
  initialize local storage $lst$\;
  \While{$BB_k$ has instruction $I_m$ not visited }{
   \If{$I_m$ is store/load instruction}{
    update $bt$\;
    obtain operand $opd$ from $I_m$\;
    \If{$opd$ is GEPOperator}{
     parse $opd$\;
    }
    update $lst$\; 
   }
  }
  update $v$\;
 }
 \caption{Memory access estimation generation.}
\end{algorithm}

\textbf{Static analysis for metric extraction}
In the design of the analyzer, we focus on efficiency and usability. Therefore, users are able to obtain all the hCFG features during compile time, while dynamic profiling is an available option to choose in case more precise data is needed. In the pure static mode, the static analyzer is utilized for collecting both the node and edge features. We build the static analyzer on top of the LLVM intermediate representation (IR), which is the bridge between lexer, parser (frontend), and code generation (backend).
In addition to better usability, when performing analysis on the IR level, it is possible for us to isolate the design of the processing logic from the types of source code to make Meliora a language-independent framework. Furthermore, it also reduces the complexity of the analysis compared to working directly on the binary code. Moreover, LLVM provides a large number of handy functions for developers to leverage. Technically, Meliora is expected to work compatibly with any programming language as long as it can be transformed into the LLVM IR. In the rest of this section, we discuss the details in the implementation of Meliora for metric extraction.

Independent of the types of the target metrics, we must inevitably traverse the abstract syntax tree (AST) or similar wrapped structure in LLVM IR one or more times, while keeping track of several values in order to summarize the corresponding metrics correctly. However, we can minimize the repeated process by adjusting the entry point of the analysis and also by aligning with the analysis granularity. Specifically, we consider loops, especially nested loops, as a whole graph so that we can flatten them from the outermost loop. After locating the top-level loop, we treat each basic block equally independent of its type (e.g. loop SCoP) and traverse once to collect necessary data from basic blocks. Each basic block is examined to retrieve and categorize the parsed instructions. For coarser-grained results, we want to collect kernel-level information which might comprise several loops at the same level. To address this problem, we first identify and locate the first and last loops then generate a fake loop body to enclose them. After that, we can reuse the same method to process the fake loop and generate a kernel graph. This approach also allows finer control over granularity within each kernel that falls between kernel-level and single-loop.

The edge features can be extracted either statically or dynamically. The dynamic method requires program instrumentation. Then the instrumented code is compiled and run to generate the hCFG transition probability edge attributes. Before Meliora comes into play, the profiled data must be merged back with the source code to create the LLVM bitcode. By contrast, in the pure static approach, we provide an LLVM component for obtaining the edge data statically, which uses heuristics to compute the edge probability based on the weights produced by the DAG analysis. We note that both the static and dynamic approaches occur before the loop traversal pass, and the edge data are collected at the same time as node attributes. However, due to the extra steps of instrumentation and execution, the time consumption for the dynamic approach is substantially higher than that of the static method. In section~\ref{sec:result} we compare the two approaches in terms of time cost and accuracy.

Describing the memory access pattern without profiling data is never a trivial task. Moreover, the enforcement of the static single assignment (SSA) form in LLVM IR complicates the implementation at the symbolic level. To address the challenges, Meliora employs the symbols extracted from the IR to estimate the bounds of the reuse distance in bytes. This is to say that we might not be able to obtain the precise memory references of an array, yet we can deduce the maximum and the minimum number of access of the same array by appropriate assumptions to compute reuse distance bounds. In the implementation, we iterate the instructions and three LLVM instruction types involve: \textit{StoreInst}, \textit{LoadInst} and \textit{GetElementPtrInst}. As shown in algorithm~\ref{tab:1}, we start by testing the instruction type and then for the qualified instructions the operand of the instruction is visited. Again the operand is tested because the instruction might manipulate the scalar values. Generally, if the operand is an LLVM $GEPOperator$ the algorithm considers the instruction as a memory access operation. Subsequently, the instruction operand is parsed to retrieve information about the array. We might need to recursively parse the operand if the target is a multi-dimensional array. The memory access estimation is generated at the same time as other metrics no additional graph traversal is required.

\subsection{Graph representation learning}\label{sec:learn}
Graph representation learning is at the heart of the Meliora workflow. The framework relies on the machine learning techniques to train the model for unseen graph prediction in order to assist the selection of the tuning parameters for code optimization. In addition to the model, it converts the raw hCFGs generated in the front end into a vector while preserving significant graph properties. The embedded form reduces the costs of storage and computation on the original graphs, which is crucial for scaling up this approach to a large number of computational patterns. A number of graph embedding options exist, as briefly discussed in Sec.~\ref{sec:background}. To choose the method most suitable to our needs, we consider both the current demand and extensibility potential. We have four requirements for any potential approach: 1) It must work for arbitrary graphs; 2) It can process auxiliary data besides the topological information of the graph; 3) It can handle both directed and undirected graphs; and 4) It can handle both vertex and edge attributes with discrete and continuous values to generate a whole-graph embedding. Relatively few approaches fulfill these requirements, leading us to the choice described below.

We build the component for graph representation learning on top of PSCN~\cite{niepert2016learning} proposed by Niepert et al. This approach is based on the convolutional neural network (CNNs)~\cite{lecun1998gradient, lecun2015deep} aiming to learn the arbitrary graph with node and edge attributes for prediction. Because our ultimate goal is to compare and match large numbers of computational kernels, we need an approach that is both \emph{accurate} and computationally \emph{efficient}. If we were to use the graph representation directly, e.g., to build a tree-based classification model, we would have to employ an expensive graph comparison operation, such as graph isopmorphism. This would limit both the size and the number of codes that can be considered. Hence, we choose to reduce the hCFG to a vector respresentation, task for which CNNs are one of the most suitable approaches. The vector representation also offers the advantage of easy comparisons using a number of different distance techniques, including cosine or Eucledian distances. In Fig.~\mbox{\ref{fig:similarity}}, for example, we show the cosine distances between the vector embeddings of test dataset loops and the embeddings of the loops in the training kernels.

The whole procedure consists of three steps. Briefly, \textit{Node sequence selection} is to construct the sequence of nodes and create the corresponding receptive fields. Followed by \textit{Neighborhood assembly} it assembles a local neighborhood for the nodes in the sequence as the candidate for the receptive field. The third step \textit{Graph normalization} is to normalize the neighborhood graph assembled in the previous step by imposing an order on the graph in order to create the vector representation. To minimize the overfitting of model training, dropout regularization \cite{JMLR:v15:srivastava14a} is applied on the hidden layer of the CNNs. The dropout rate is set to 50\%. The procedure also trains a model utilized by Meliora to predict unseen graphs.

\subsection{Using the Model}\label{sec:usingmodel}

After the model generation, users can apply Meliora to key loops in their code to locate the best match for an arbitrary graph with the kernels in the model. To achieve that, first Meliora compiles the source code in any language supported by LLVM into bitcode and then performs the static analysis described in Sec.~\ref{sec:analysis} on the bitcode to collect the hCFGs representing the loops of the target kernel. One graph is created for each loop in the loop-level mode, otherwise a single graph for the entire kernel is generated. Subsequently, Meliora uses the model to make predictions consisting of the coefficient vector containing the probabilities of the given graph being similar to kernels used in training. We then choose from this vector the loop or kernel with the largest coefficient, i.e., the best match. At present, this is where Meliora stops, but in the future, we plan to integrate it more closely into the autotuning process. We note that this approach can be used for different types of optimization workflows, not just autotuning. As our training data grows, we anticipate that we would incorporate both manually optimized and autotuned code versions, so depending on the specific match, the user may embark on a manual optimization, code replacement (if a better implementation of the same functionality was located), or enlist the help of an autotuner.  We include some specific examples on a possible autotuning integration workflow in the evaluation Section~\ref{sec:eval-newcode}.

%% file: evaluation.tex
\section{Evaluation}\label{sec:result}

In this section, we evaluate the accuracy of the model and the performance of the process of model generation. The datasets we used to build the model are from the SPAPT~\cite{balaprakash2012spapt} benchmark.

\subsection{Experiment environment}\label{sec:env}

The machine we used to build the model and validate it is an \intelxeon E5-2699v3 2.30GHz with two 18-core Haswell CPUs and 256GB of memory.

\subsection{Dataset Generation}\label{sec:training}
It is not easy to collect a sufficient amount of code as the input for training. Hence we create our dataset from scratch for the selected kernels. As we mentioned, Meliora is capable of running in a postmortem mode to work in conjunction with the Orio autotuning framework~\cite{Norris:2007,hartono2009annotation}. In that scenario, Meliora serves as a post-processor invoked by the autotuner to perform the static analysis on the various versions of the tuned code generated by Orio. No modifications to Orio were necessary; we believe integration with other autotuners can be accomplished similarly.

\begin{lstlisting}[language=c,caption={Annotated AXPY-4 kernel}, frame=bt, belowcaptionskip=0.8em, label={list:2},basicstyle=\fontsize{8}{10}\ttfamily, commentstyle=\fontsize{8}{10}\ttfamily\itshape\color{blue}]
  // Tuning specification
  /*@ begin PerfTuning (
    // ...
    def performance_counter {
     arg method = 'basic timer';
     arg repetitions = 60;
    }
    def performance_params {
     param UF[] = range(1,33);
    }
    //  ...omitted ...
    def search {
     arg algorithm = 'Randomsearch';
    }
 ) @*/
  /*@ begin Loop ( 
    transform Unroll(ufactor=UF) 
    for (i=0; i<=N-1; i++)
        y[i] = y[i] + a1*x1[i] + a2*x2[i] 
                + a3*x3[i] + a4*x4[i];
  ) @*/
  // original C code kernel
  for (i=0; i<=N-1; i++)
        y[i] = y[i] + a1*x1[i] + a2*x2[i] 
                + a3*x3[i] + a4*x4[i];
  /*@ end @*/
  /*@ end @*/

\end{lstlisting}

The Orio autotuning framework parses the annotations in the source code and generates different versions of the optimized code. Next, Orio empirically evaluates all the generated versions to select the one with the best performance for the production environment. Listing~\ref{list:2} shows a portion of the annotated AXPY-4 kernel. In the example, the \textit{performance\_params} defines only one performance parameter the unroll factor (UF) ranging from 1 to 33.  Orio will employ a random search strategy (as specified in the tuning spec comment) to determine empirically the optimal unrolling factor for this loop on the platform of interest. \textit{input\_params} defines two different problem sizes that the entire tuning process will repeat for each value in order to search for the best performance for each size.

\begin{table}[h!tb]
\centering
\caption{Selected kernels for dataset generation}
\label{tab:2}
{\renewcommand{\arraystretch}{1.2}
\begin{tabular}{|c|c|}
\hline
\multicolumn{1}{|c|}{\textbf{Kernel}}     
    & \multicolumn{1}{|c|}{\textbf{Operation}}  \\ \hline\hline
    \multicolumn{2}{|c|}{Elementary linear algebra kernels} \\ \hline
GEMVER        & scalar, vector, and matrix multiplication \\  
MVT           & matrix vector product and transpose     \\ \hline\hline
\multicolumn{2}{|c|}{Stencil code kernel} \\ \hline
Stencil3d     & 3D stencil computation \\ \hline\hline
\multicolumn{2}{|c|}{Linear solver kernel} \\ \hline
BiCG          & subkernel of BiCGStab linear solver \\ \hline\hline
\multicolumn{2}{|c|}{Elementary statistical computing kernel} \\ \hline
COV           & covariance computation \\ \hline

\end{tabular}}
\end{table}

The dataset we used for training is a portion of the SPAPT benchmark suite~\cite{balaprakash2012spapt,SPAPT}, which contains four types of selected kernels. Table~\ref{tab:2} shows the kernel name and major operations grouped by the categories. 1) \textit{Elementary linear algebra kernels} focus on the mathematical computations on scalars, vectors, and matrices. 2) The \textit{Stencil code kernel} is frequently used for solving partial differential equations. They follow a particular pattern to access and modify the array elements. 3) The \textit{Linear solver kernels} are used in solving systems of linear equations. For instance, the \textbf{BiCGStab} linear solver kernel decomposes a matrix into a product of lower and upper triangular matrices. 4) \textit{Elementary statistical computing kernels} refer to the code that helps find the statistical relationship among random variables. The dataset used in the evaluation is listed as the following. Each of them is split into two groups for training (approximately 3/4) and validation (approximately 1/4), respectively.
The dataset containing loop versions is statically generated and is divided into two subsets: 5,201 graphs for modeling and 1,498 graphs for validation. The test graphs are versions of the same kernels and can be used to confirm that the matching is able to identify such known similar codes correctly (Figures~\ref{fig:val_a} and~\ref{fig:val_b} show the self-validation at kernel- and loop-level granularity, respectively; in addition, we validated with codes not used in training~\mbox{\ref{fig:similarity}}, as discussed later).

\subsection{Results}
In this section, we discuss the accuracy of Meliora in identifying the kernels which we split into two parts: (i) the accuracy for recognizing different versions of kernels used in training (Sec.~\ref{sec:seen}) and (ii) effectiveness of using Meliora to optimize completely arbitrary (unseen) code (Sec.~\ref{sec:eval-newcode}). 

\subsubsection{Model Validation}\label{sec:seen}
As we described in Sec.~\ref{sec:training}, we split a subset of the kernels (Table~\ref{tab:2}) in the SPAPT benchmark suite into two groups,  using the majority (3/4) for the model training, while the rest of the code versions (1/4) are reserved for model validation. In this scenario, we are able to test the accuracy of the model for recognizing different versions of the same kernels used in training. 

\begin{figure}[h!tb]
    \includegraphics[width=\linewidth]{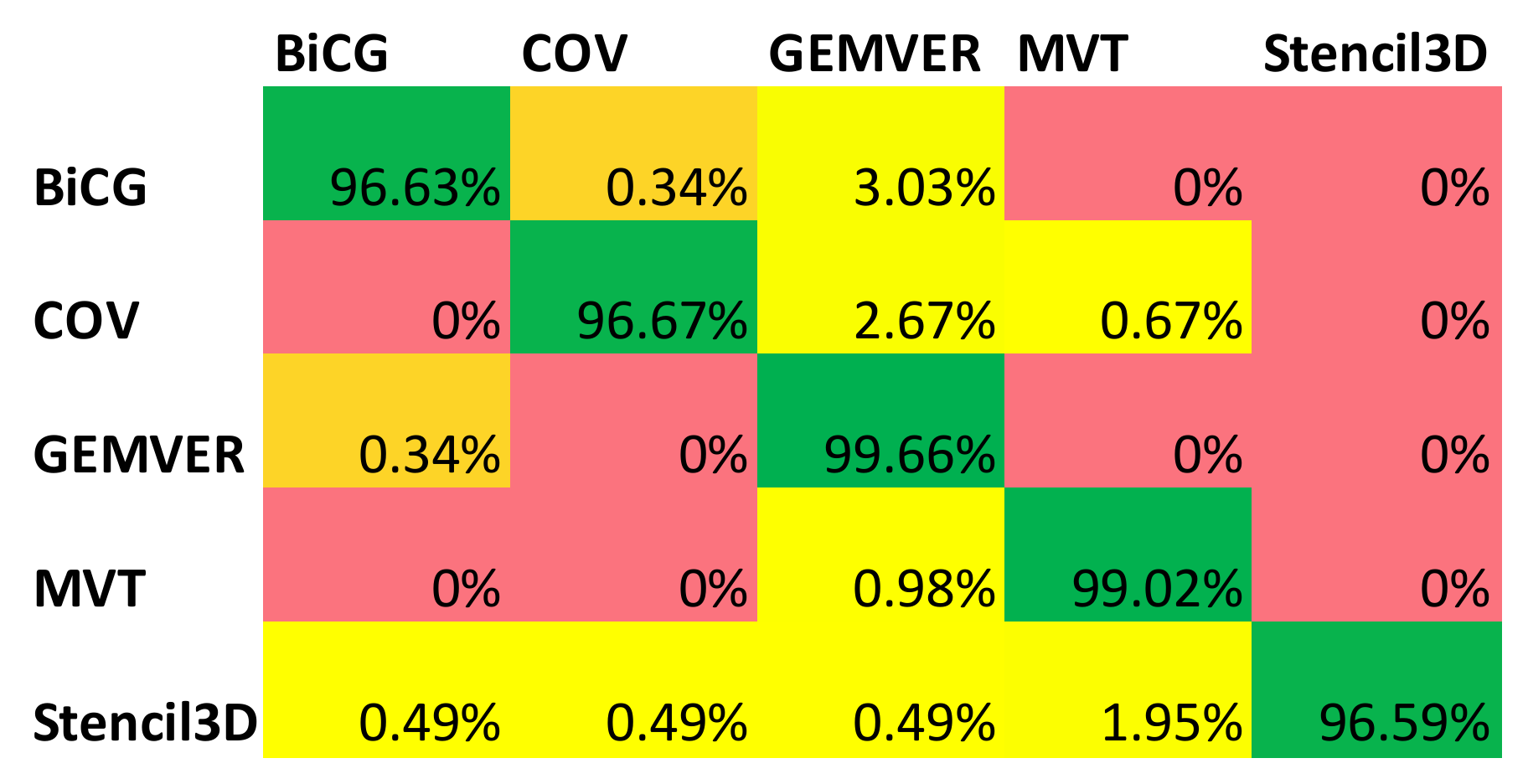}
\vspace{-10pt}
\caption{Model self-validation at kernel-level granularity.}
    \label{fig:val_a}
\end{figure}

Figures~\ref{fig:val_a} shows the self-validation results on the five-kernel dataset used for training. By self-validation, we mean selecting a transformed (by Orio) kernel version that was not used in training, and computing its match; for example, we expect that most versions of GEMVER would be matched with other versions of GEMVER. This is not a completely trivial validation since many of the transformations impact the hCFG and to a lesser extent, the instruction mix. The labels on the X-axis and Y-axis are the same as the kernel names where the X-axis represents all the available classes in the training set corresponding to each of the selected kernels, and the Y-axis indicates the percentage of the graphs in the validation set predicted as the existing kernels. The color of the tiles shows the value of the percentage, the redder the tile is the closer the value to 0 and similarly, greener means the value is closer to 100.

From Figure~\ref{fig:val_a} we can see that most of the cells on the diagonal are colored dark green which shows that the majority of the graphs are correctly predicted. In particular, the validation results are all above 96\% ranging from 96.59\% to 99.66\%. 
On the other hand, the percentage of misclassified graphs is less than 4\%. MVT shows the best accuracy with only 0.98\% of the graphs incorrectly labeled as GEMVER; Graphs from Stencil3D are the most misclassified results in the 3.41\% of the graphs classified as each of the other training classes. In practice, for a given graph we perform prediction and then choose the most likely matching result based on the prediction vector which includes the probability of a likely matching against each class in the model. In other words, we always choose the kernel with the highest value.

\subsubsection{Evaluation on New Kernels}\label{sec:eval-newcode}
Most real-world use cases of Meliora would utilize the model to find the best match between unknown, arbitrary kernels and those in the model so that we can apply existing optimization knowledge to avoid the time-consuming search on the large variant spaces of performance optimizations. 
To demonstrate the application of Meliora to new codes, we present results of using it on a subset of the SPAPT benchmarks that \emph{were not} used at all for building the model.

We performed the empirical evaluation on a slightly different Haswell system than the server used in the model generation (for scaling and cost reasons); the system we used was a cluster of \intelxeon CPU E5-2690 v4 @ 2.60GHz nodes. Each node has two 14-core CPUs and 128GB of memory. 

The evaluation procedure consists of the following steps. First, we apply Meliora as described in Sec.~\ref{sec:usingmodel} to the subset of SPAPT codes shown in Table~\ref{tab:newcodes}. None of these computations were used for training. The codes typically contain several loops with various nesting depths and range in size from a tens to hundreds of lines of C code. 

\begin{table}[h!tb]
    \centering
\caption{Meliora's matches for a set of new codes' loops.}
    \label{tab:newcodes}
    \begin{tabular}{|l|l|c|}\hline
\textbf{New code} & \textbf{Matched kernel} & \textbf{Coeff.}\\
\textbf{Name@LoopLoc.}        &                       & ($<=1$)\\ \hline
adi@132 & 	gemver &	0.99 \\
adi@137 & 	gemver &	0.99\\
correlation@166 &	covariance &	0.99 \\
correlation@172	 & mvt &	0.99 \\
correlation@180 &	mvt &	0.85 \\
correlation@185	 & covariance &	1.00 \\
fdtd@152 &	bicgkernel &	0.99 \\
fdtd@154 &	mvt &	0.88 \\
fdtd@157 &	mvt &	0.88 \\
fdtd@160 &	mvt &	0.83 \\
jacobi@76 &	gemver &	0.99 \\
tensor@130 &	stencil3d &	1.00 \\
trmm@124 &	covariance &	0.99 \\
trmm@130 &	stencil3d &	1.00 \\
dgemv@255 &	bicgkernel &    0.83 \\
dgemv@258 &	gemver &    1.00 \\
dgemv@260 &	bicgkernel &    0.83 \\
dgemv@263 &	gemver &    1.00 \\
dgemv@265 &	bicgkernel &    0.91 \\
dgemv@268 &	gemver &    1.00 \\
dgemv@270 &	bicgkernel &    0.99 \\
dgemv@276 &	gemver &    1.00 \\
dgemv@278 &	gemver &    1.00 \\
dgemv@280 &	gemver &    1.00 \\\hline
    \end{tabular}
\end{table}

\begin{figure}[h!tb]
\includegraphics[width=\linewidth]{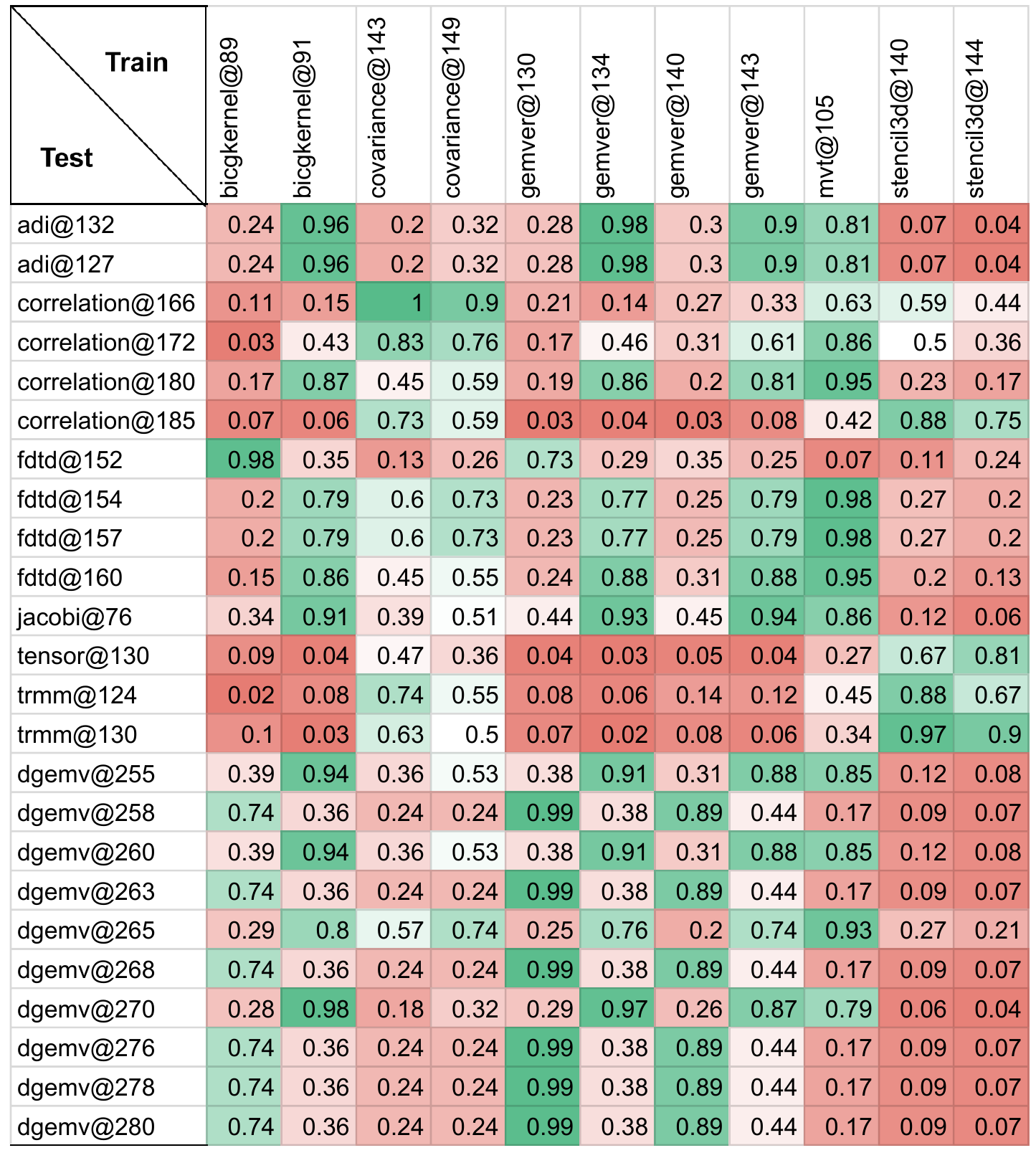}
\caption{Similarity between test kernels and training dataset loops computed as the cosine distance between their hCFG embeddings.}
\label{fig:similarity}
\end{figure}

Once the model returns a match, we compute the similarity score (cosine similarity) between the loops in the new kernel and those the matched kernel to further refine the mapping. For example the first loop in adi (adi@132) is matched with the gemver kernel, which has four loops. The cosine similarity between the embeddings of the adi@132 and the gemver@134 loops is the highest, hence we finalize the match to be adi@132-gemver@134. For small datasets, one could just compute similarities instead of employing the kernel-level model, however, as the number of kernels in the training dataset grows, this approach would not scale.

Next, for each of the target loops (indicated by their line number in the left column of the table), we manually transfer the optimal tuning parameters from the corresponding loop in the \emph{matched kernel} (middle column). Because we used Orio to create our training data set, we also simultaneously produced autotuned versions of these kernels. Given the kernel name, problem sizes, and architecture, we can then easily look up the set of transformations that produced the best version of that kernel for that problem size and architecture. Figure~\ref{fig:similarity} shows the cosine similarity between the vector representations of the training and test datasets. We show similarities between the original code versions for each training and test kernel.

The next step in our evaluation is to manually copy the tuning spec from the autotuned version of the matched kernel into the new code, adjusting variable names as needed. For example, the \textbf{correlation} code has four loops, two of which were matched with \textbf{covariance}, and two with \textbf{mvt}. In future, we plan to significantly automate this step, while still allowing the user to customize the inserted tuning annotations.

Because the matched kernels often have more than one potential loop, the user either must select the loop from which to copy the tuning options or combine all possible parameters. The first choice eliminates empirical autotuning altogether and is often possible through quick inspection of both codes, code similarities are often obvious to a human once presented with such limited choices. In a few cases, it is difficult to choose just one loop; in these situations, parameters from multiple loops can be combined, and the new code can be empirically autotuned. While this does not completely avoid the cost of empirical tuning, it does reduce it dramatically (we give some more details later in this section). Individual loop matching can potentially be fully automated, although it is not clear yet whether the explosion in complexity is warranted to save some fairly light manual effort---all the results in this section were produced in a few hours, including creating the tuning specs for the new codes and performing the autotuning for some of the kernels. 

\begin{table}[h!tb]
    \centering
\caption{Search speedup for the unseen kernels.}
    \label{tab:searchsu}
    \begin{tabular}{|l|c|}\hline
\textbf{Kernel} & \textbf{Search speedup} \\\hline
adi & 	19 \\
correlation & 1126 \\
dgemv &  2 \\
fdtd	 & 19 \\
jacobi & 10 \\
tensor & 14 \\
trmm  & 17 \\\hline
    \end{tabular}
\end{table}

\begin{figure}[h!tb]
  \centering
    \includegraphics[width=.9\linewidth]{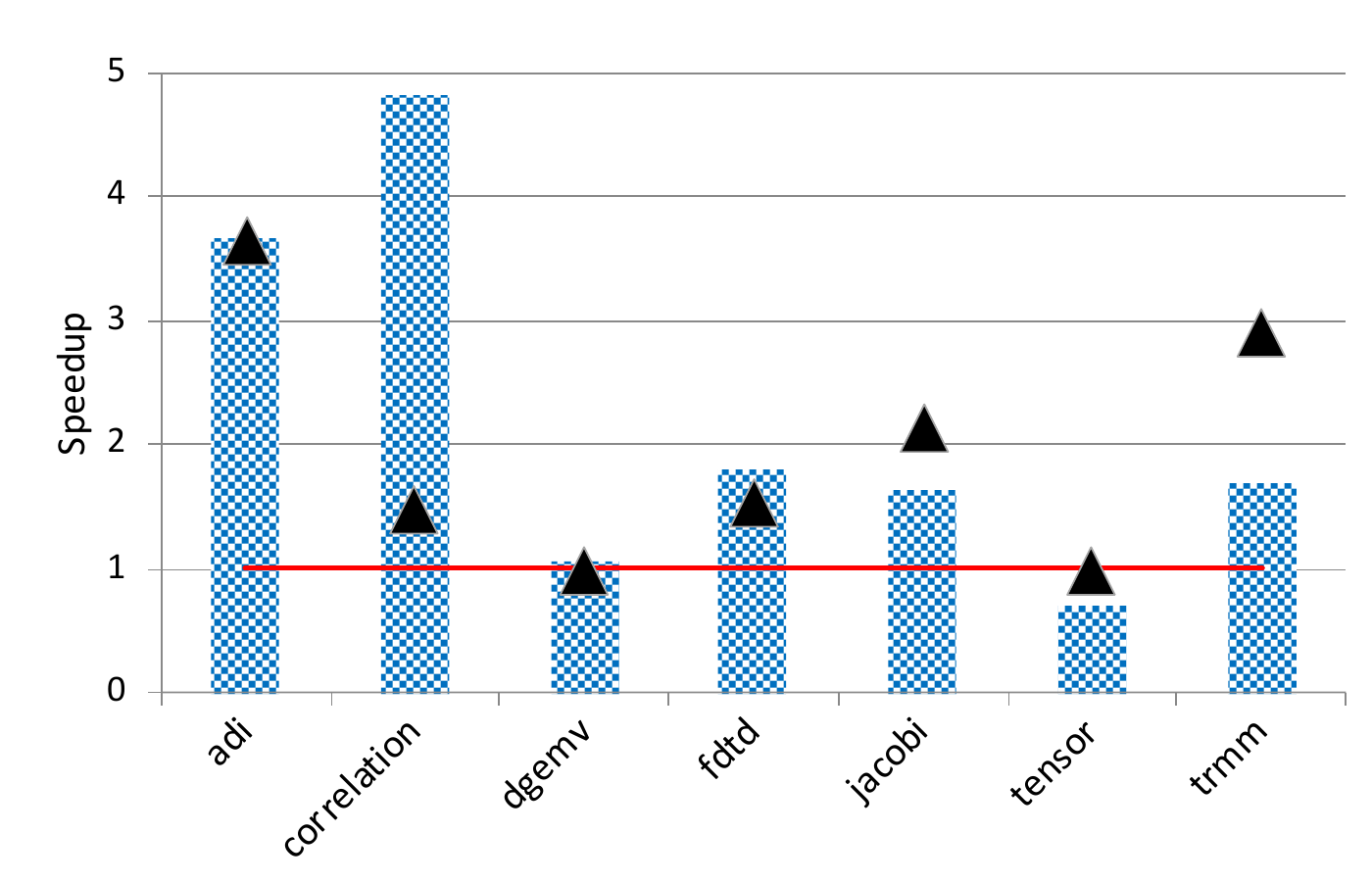}
    \vspace{-10pt}
  \caption{Speedup over the unoptimized (base) versions: bars for Meliora-matched optimizations, black triangles for empirical autotuning results, and a red line for the baseline performance.}
  \label{fig:speedup}
\end{figure}

\paragraph{Kernel performance} The speedups obtained by modifying the new codes as described above are shown in Fig.~\ref{fig:speedup}. The baseline is the original, unoptimized version, compiled with a recent GCC compiler using the -O3 optimization level. All optimization options in the matched kernels can be seen in the SPAPT benchmark repository~\cite{balaprakash2012spapt,SPAPT} and include loop unrolling, cache tiling, register tiling, SIMD pragma insertion, OpenMP parallelization, and scalar replacement. While we used an autotuner to enable rapid application of these optimizations, one could also apply them manually, albeit at a dramatically increased effort (the size of the tuned code is typically much larger than the original, especially when combining multiple transformations). 

We completely eliminated the empirical search \emph{and} produced a better-performing version for some of the new codes (\textbf{adi}(1.78x\footnote{Speedup with respect to the original code version.}), \textbf{correlation}(4.2x), and \textbf{trmm}(1.12x)). In addition, we were able to improve performance further by applying limited autotuning for \textbf{adi} (3.7x), \textbf{dgemv} (1.1x), \textbf{fdtd} (1.8x), \textbf{jacobi} (1.6x) and \textbf{correlation} (4.8x); this required minimal extra effort to modify the parameter space to include the default options. In general, as Table~\ref{tab:searchsu} shows, we reduced the search time by 2x to 1,126x for the test kernels, with identical search method configurations.

\begin{figure}[h!tb]
  \centering
    \includegraphics[width=\linewidth]{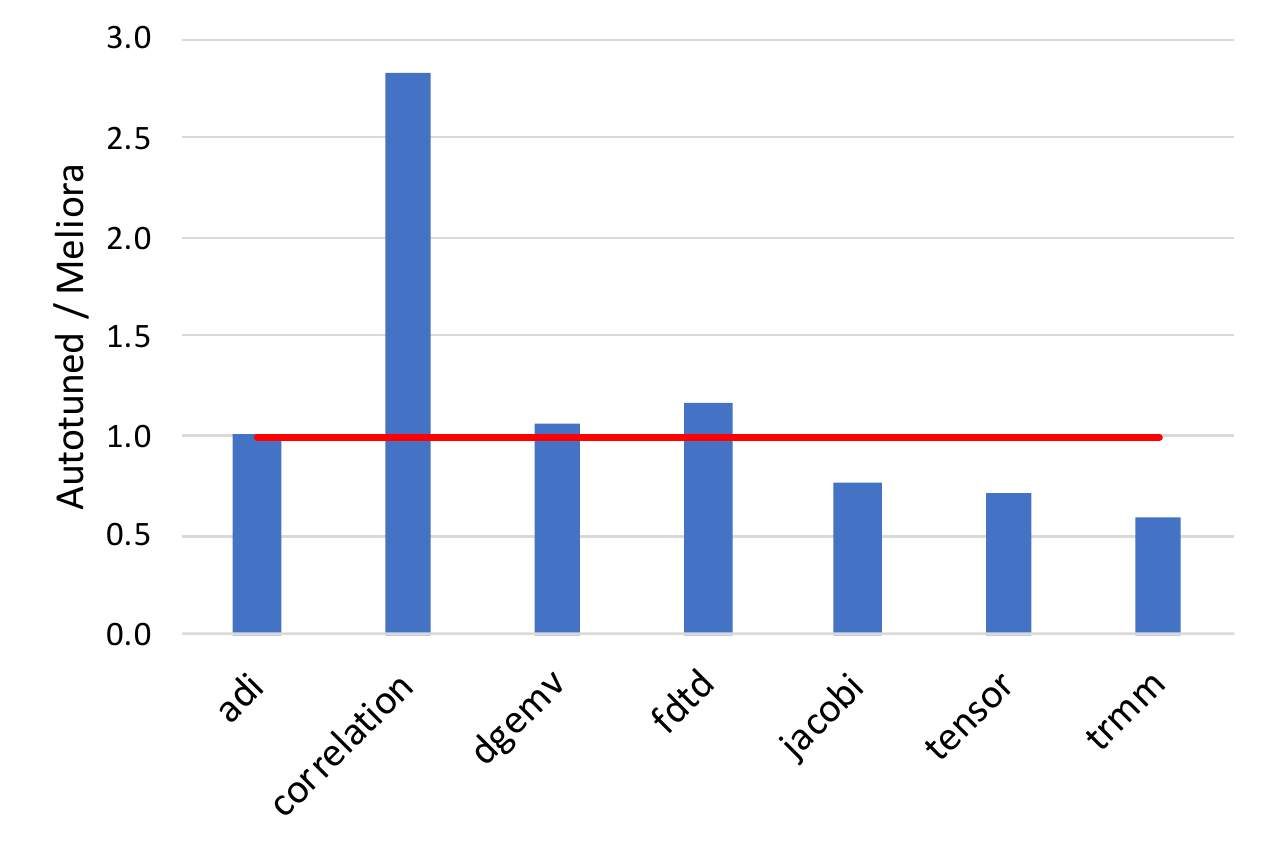}
    \vspace{-20pt}
  \caption{Comparison with empirically tuned performance. Values greater than one indicate that Meliora-based optimizations outperformed the empirical autotuner.}
  \label{fig:compare_autotuned}
\end{figure}

Figure~\ref{fig:compare_autotuned} shows a comparison between Meliora-based optimized codes and empirically autotuned versions using a machine-learning-based search strategy capped at 1000 runs (the same as was used to generate the model, although in most cases, fewer than 100 runs were performed per kernel by the search method). Such capping is necessary because the size of the parameter search spaces (ranging from $10^4$ to $10^24$ for these codes) is too large to allow exhaustive search. The only code for which the autotuner significantly outperformed the Meliora-based version is \textbf{trmm}. For \textbf{correlation}, using the tuning specification from the matched kernels significantly outperformed the result from autotuning the original by providing a better starting point for the search, as well as a slightly different set of optimizations.

\paragraph{Autotuning search performance} We evaluated performance both on a single set of parameters and with limited autotuning on a small parameter space based on the matched loops. The post-match autotunning also benefits from the use of local optimization methods because we know that the starting point is likely close to the optimum.

For the \textbf{correlation} benchmark, the Meliora-based optimization through a match with loops from the autotuned \textbf{covariance} and \textbf{mvt} kernels), we were able to actually outperform previous empirical tuning without any autotuning. The results for \textbf{tensor} indicate that this specific benchmark is not optimizable via the kinds of optimizations we attempted (as indicated by the fact that both the Meliora and autotune results are close to the original performance). In part this is due to the fact that it contains a five-level loop; there is nothing similar among the other kernels in SPAPT. To use Meliora for such cases, a model should be trained with a greater number of representative kernels, including tensor contractions.

\begin{figure}[h!tb]
  \centering
    \includegraphics[width=0.46\textwidth]{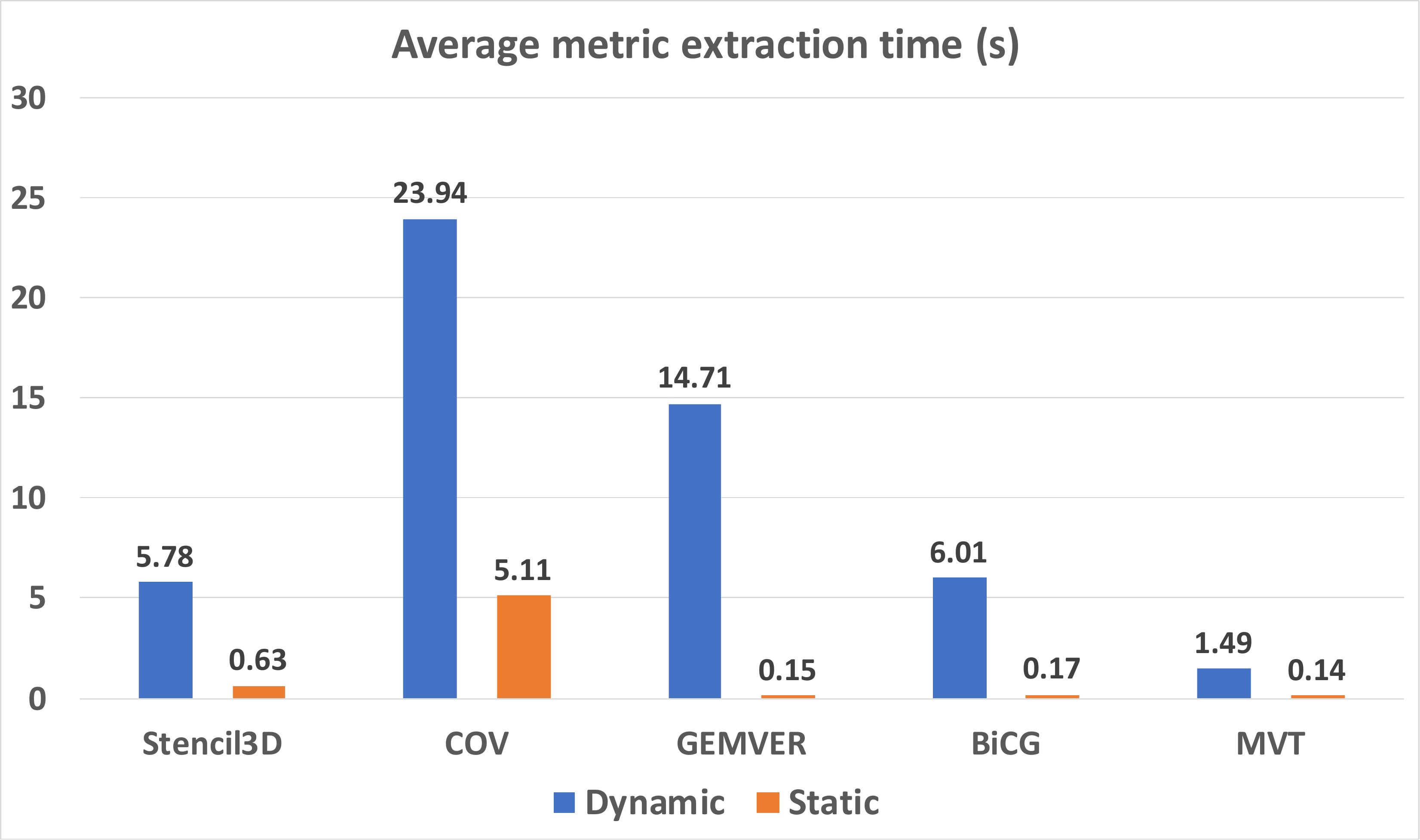}
  \caption{Comparison between the static and dynamic strategies for metric extraction. The X-axis lists the kernels and the Y-axis is the time in seconds. }
  \label{fig:time}
\end{figure}

\paragraph{Performance of metric extraction}
In addition to the model validation, we time the process of loop-level metric extraction illustrated in Figure~\ref{fig:time}. The timed process is one component in the workflow of Meliora to generate hCFGs. 
The plot shows the average time spent on source code compilation (to LLVM IR) and conducting the information retrieval. 

The metric extraction time varies from kernel to kernel depending on the size of the generated code, which is determined by both the kernel and the input parameters set by Orio annotations. Since the dynamic method demands additional steps to prepare and run the instrumented code, it is not surprising that it takes longer to collect the metrics. Among the kernels, COV (\textbf{covariance}) is the most time-consuming and it needs approximately 5 seconds on average even for the static processing of the graphs, which is mainly spent on the compilation and I/O operations for the relatively large graphs.

%% file: related.tex
\section{Related Work}\label{sec:related}
In this section, we briefly survey some works to the application of graph representation learning to performance modeling and model-based performance optimization. 

The area of compiler optimization and autotuning shares several concepts in common. For instance iterative compilation~\cite{bodin1998iterative, kisuki2000combined}  exhaustively searches the parameter space for the best-performing combinations. However, compilers usually work at a broad level which make more efforts to generalize the optimal variants into a heuristic to optimize all codes, while autotuners usually focus on specific applications. Analytical models derived manually are used at the early stage of compiler optimization. Wagner et al.~\cite{wagner1994accurate} and Tiwari et al.~\cite{tiwari1994power} hand-tuned the analytical model to estimate the execution frequency of code regions and the energy consumption. Nevertheless, creating the model by hand requires expertise and significant effort, and the accuracy and the portability of the model completely depends on the human factors.

Machine learning alleviates the complexity of analytical model creation and has been a technique widely used in compiler optimization. 
Wen et al.~\cite{wen2014smart} and Luk~\cite{luk2009qilin} leverages the regression on predicting the application speedup and running time. Classification is also used for locating the optimal optimization parameters~\cite{stephenson2005predicting, yuki2010automatic, beckingsale2017apollo}. In addition, clustering as an unsupervised learning method is used to select the optimal execution point for program simulation~\cite{perelman2003using}. Comparing with the surveyed works, our proposed method concentrates on the intrinsic features of the code as the complementary attributes to the topology information and also emphasizes the automatic feature extraction in order to further reduce the human efforts involved and improve the framework efficiency.

Although our research goals are different, we conduct research motivated by the same ideas which leverages graph representation learning for similarity detection. Thus it is valuable to mention other such works in this section. For example Xu et al.~\cite{xu2017neural} and Liu~\cite{liu2018alphadiff} utilize learned embedded format of the control-flow graph extracted in binary functions to compute the distance in order to measure the binary similarity. They consider only the structure and do not include instruction mix information. Lim et al.~\cite{lim2018similarity} define CFG-based similarity metrics for matching GPU kernel computations that also include instruction mix information; that approach is limited to NVIDIA codes, but integration with our CPU approach is a feasible future direction.

%% file: conclusion.tex
\section{Conclusions and Future Work}\label{sec:future}
In this paper, we introduce Meliora, a framework for extracting code representations that can be used to find potential optimizations for new codes more easily and eventually, automatically. 
The model validation and use cases we consider suggest that the metrics we defined can indeed produce an accurate model of loop-based computations, and showed how we can eliminate or greatly reduce code optimization efforts with or without autotuning while achieving competitive results compared to traditional empirical autotuning. 

However, much work remains to be done beyond this initial proof of concept. Our future work includes identifying more features on which to base metrics that are more representative \emph{and} low cost in extraction. For example, a few input parameters (typically related to problem size) can have a significant impact on optimizations and it may be a good idea to somehow flexibly and generally incorporate them into our model. We plan to extend Meliora with inter-procedural analysis. We are also investigating new metrics on the extended hCFG to enable Meliora to perform the analysis on parallel applications.
In addition, we plan to complete the integration of Meliora into the autotuning process for speeding up autotuning of new codes by reducing the search space.